# From Textbook to Talkbot: A Case Study of a Greek-Language RAG-Based Chatbot in Higher Education

Maria Eleni Koutsiaki, International Hellenic University, Greece
Marina Delianidi, International Hellenic University, Greece
Chaido Mizeli, International Hellenic University, Greece
Konstantinos Diamantaras, International Hellenic University, Greece
Iraklis Grigoropoulos, International Hellenic University, Greece
Nikolaos Koutlianos, Aristotle University of Thessaloniki, Greece

**Abstract**
The integration of AI chatbots into educational settings has opened new pathways for transforming teaching and learning, offering enhanced support to both educators and learners. This study investigates the design and application of an AI chatbot as an educational tool in higher education. The chatbot was developed utilizing instructional materials from the course Family Psychology offered by the Department of Early Childhood Education and Care at the International Hellenic University, in conjunction with curricular resources from Sports Medicine courses by the School of Physical Education and Sport Science at the Aristotle University of Thessaloniki. Designed to operate in the Greek language, the chatbot addresses linguistic challenges unique to Greek while delivering accurate, context-grounded support aligned with the curriculum. The AI chatbot is built on the Retrieval Augmented Generation (RAG) framework by grounding its responses in specific course content. RAG architecture significantly enhances the chatbot's reliability by providing accurate, context-aware responses while mitigating common challenges associated with large language models (LLMs), such as hallucinations and misinformation. The AI chatbot serves a dual purpose: it enables students to access accurate, on-demand academic support, and assists educators in the rapid creation of relevant educational materials. This dual functionality promotes learner autonomy and streamlines the instructional design process. The study aims to evaluate the effectiveness, reliability, and perceived usability of RAG-based chatbots in higher education, exploring their potential to enhance educational practices and outcomes as well as supporting the broader adoption of AI technologies in language-specific educational contexts. Findings from this research are expected to contribute to the emerging field of AI-driven education by demonstrating how intelligent systems can be effectively aligned with pedagogical goals.

**Introduction**

Artificial intelligence (AI) has demonstrated significant potential for application across diverse domains of knowledge. In contemporary education, AI can serve as a flexible tool for designing and developing educational material, enhancing learner engagement through interactive experiences, and supporting the assessment and evaluation of student performance (Wang et al., 2024). Among the various real-world applications of AI, digital assistants—commonly referred to as chatbots—have emerged as one of the most widely adopted and practically impactful forms.

The integration of AI-powered chatbots into educational contexts has introduced new opportunities for transforming teaching and learning practices. These systems provide enhanced support for both educators and learners, enabling more personalized, responsive, and efficient educational experiences. According to Gupta (2020), chatbots can be classified into three primary categories based on their underlying technologies, employed algorithms, and user interface design:
1. **Menu/Button-Based Chatbots**, which offer users a predefined set of options and direct them to responses corresponding to their selections.
2. **Keyword Recognition-Based Chatbots**, which identify specific words or phrases within a user's input and generate relevant responses accordingly.
3. **Contextual Chatbots**, which interpret not only the literal meaning of a user's message but also their underlying intent, purpose, or emotional state. These chatbots maintain conversational histories, allowing them to learn continuously and refine their interactions over time. As a result, their responses increasingly approximate natural human communication.

The incorporation of chatbots into educational environments has proven to be particularly beneficial, offering diverse applications that support both teaching and learning processes. As noted by Kasneci et al. (2023), chatbots are utilized across multiple educational levels—including primary, secondary, and higher education—as well as in special education, lifelong learning, and vocational training.

As pedagogical tools, chatbots offer a broad spectrum of functionalities, as discussed by Kasneci et al. (2023) and Wang et al. (2024). They can act as digital assistants to improve reading and writing proficiency through automated correction of spelling, grammatical, and syntactical errors. They serve as always-available learning companions, providing immediate assistance and continuous feedback. Moreover, chatbots can foster critical thinking by generating questions and problem-solving tasks, facilitate practice through quiz generation, and assist learners in organizing and managing their academic workload. Additionally, they hold particular promise for supporting learners with disabilities by delivering adaptive written or spoken language responses and personalized instructional assistance.

This study examines the design and implementation of an AI chatbot developed as an educational tool within higher education contexts. The chatbot is built upon the Retrieval-Augmented Generation (RAG) framework, which grounds the system's responses in domain-specific course materials. By integrating retrieval-based components with generative language modeling, the RAG architecture enhances response reliability and contextual accuracy while mitigating common challenges associated with Large Language Models (LLMs), such as hallucinations and misinformation.

The RAG framework optimizes the output of pre-trained LLMs by incorporating external knowledge sources through document retrieval. This process enables the model to extend its internal knowledge base and generate more accurate, contextually grounded responses (Lewis et al., 2020). Specifically, by dynamically retrieving relevant information, the RAG architecture effectively reduces the limitations of LLMs and minimizes the probability of generating false or unsupported statements (Arslan et al., 2024).

The architecture comprises three core components: information retrieval, response generation, and model optimization. Each RAG system includes:
a) the Retriever, which accepts user queries, converts them into vector embeddings, and identifies the most relevant documents or text passages within a database based on similarity metrics, and
b) the Generator, typically a sequence-to-sequence transformer model (e.g., BART - Bidirectional and Auto-Regressive Transformers or T5 - Text-to-Text Transfer Transformer), which synthesizes responses by integrating the retrieved passages with the user's query.

A critical element of the RAG workflow is embedding creation and indexing, where documents are segmented into smaller textual units (chunks) and transformed into high-dimensional vector embeddings stored in a vector database. During inference, the user's query is similarly embedded and compared to indexed chunks to retrieve semantically related content. Importantly, RAG-based systems can undergo continuous learning, allowing iterative improvement of both the retriever and generator components (Dimitrova, 2025; Lewis et al., 2020).

Implementing such systems in the Greek language presents additional Natural Language Processing (NLP) challenges due to the language's rich morphology, flexible syntactic structure, and inflectional complexity (Papantoniou and Tzitzikas, 2020). These linguistic characteristics complicate text segmentation and chunking processes compared to English, underscoring the need for tailored approaches adapted to the Greek linguistic context.

Accordingly, this study seeks to evaluate the effectiveness, reliability, and perceived usability of RAG-based chatbots in higher education, focusing on how different chunking methods and retrieval depths (Top-K values) affect system performance by setting three evaluation criteria: content accuracy, completeness, clarity. The research explores these parameters across two disciplinary domains—Family Psychology and Sports Medicine—to assess how varying content complexity and corpus characteristics influence response quality.

The findings aim to contribute to the emerging field of AI-enhanced education by demonstrating how intelligent retrieval-augmented systems can support teaching and learning processes, particularly in linguistically complex environments. The research questions guiding this investigation are as follows:
RQ1. Which chunking model among fixed-size και delimiter-aware Chunking Methods for RAG achieves the highest overall performance within the RAG-based chatbot?
RQ2. For each model, which of the three evaluation criteria (content accuracy, completeness, clarity) represents its strongest aspect?

**Methodology**

*Chatbot Design and RAG Architecture*

The primary objective of the chatbot developed in this study was to investigate the application of AI-powered conversational agents in higher education contexts. Specifically, the system was implemented using a RAG architecture designed to ground the chatbot's responses in course-specific content. This implementation also facilitated the evaluation of various text chunking strategies to identify the configuration that yields the most accurate and complete responses based on three evaluation criteria:

1. **Content Accuracy:** Assesses whether the chatbot's response directly addresses the user's query, provides factually correct and contextually grounded information, and avoids factual or logical errors.
2. **Completeness:** Evaluates the extent to which the response fully covers all aspects of the query without omitting relevant information.
3. **Clarity:** Examines the coherence and readability of the response, including the logical organization of ideas and ease of comprehension for the user.

For the purposes of this study, six distinct chunking strategies were implemented to segment the input text into discrete units ("chunks"). The effectiveness of each strategy was assessed to determine its impact on response quality according to the aforementioned criteria. The method employed for text chunking is a critical determinant of the overall performance of RAG-based systems, as it directly influences both the retrieval accuracy and the contextual coherence of the generated responses. Poorly designed chunking approaches can therefore degrade system functionality by impeding the efficient linkage between information retrieval and response generation.The chatbot system followed the standard RAG pipeline architecture, comprising two main components: a **retriever** and a **generator**.

For the retrieval stage, a vector-based semantic search method utilizing cosine similarity was employed to identify the most relevant text chunks. Document embeddings were generated using the *"dimitriz/st-greek-media-bert-base-uncased"* sentence transformer model, selected for its proficiency in processing the Greek language and capturing semantic relationships in Greek academic texts. The retrieved text segments were then provided as input to the generator component, implemented using the **Gemini Flash 2.0** large language model. This model synthesized the final responses by integrating the user query with the retrieved passages, ensuring both semantic coherence and contextual fidelity.

This architecture enabled a controlled and consistent evaluation of the different chunking strategies under uniform retrieval and generation conditions, thereby providing a robust framework for assessing the impact of text segmentation methods on RAG-based chatbot performance.

*Chunking Strategies (Six Models)*

As mentioned above, we implemented six distinct chunking strategies, grouped into two functions - character-based splitting and paragraph-based splitting.
The first applied function *(split_text_into_chunks)* splits the input text into chunks based on a specified maximum length of characters. Its parameters include the *input text* and the maximum length limit *max_len=[parameter value]*. The function preserves sentence integrity, ensuring that sentences are not split into chunks. The output is a list of text segments - chunks. This function was applied to four variants (Models 1–4), each corresponding to a different *max_len parameter value* = 200, 400, 500, and 800. These configurations produced different chunks sizes.

The second case involved dividing the text into paragraphs with two options:
1. using the function *split_text_into_paragraphs_using_sentences(text)* - Model 5, which first splits the text into individual sentences and then groups them into paragraphs by combining a fixed number of sentences per paragraph.
2. using the function *split_text_into_paragraphs_newline(text)* - Model 6, which preserves the original paragraph structure by using blank line separators as boundaries. Given the varying lengths of paragraphs, this method maintains the natural diversity found in the source text.

Table 1. Chunking splitting strategies

| Model | Chunking splitting strategy | Parameter |
|---|---|---|
| **Model 1** | *split_text_into_chunks(text, max_len=[parameter value])* | 200 |
| **Model 2** | *split_text_into_chunks(text, max_len=[parameter value])* | 400 |
| **Model 3** | *split_text_into_chunks(text, max_len=[parameter value])* | 500 |
| **Model 4** | *split_text_into_chunks(text, max_len=[parameter value])* | 800 |
| **Model 5** | *split_text_into_paragraphs_using_sentences(text)* | 3 sentences in a paragraph |
| **Model 6** | *split_text_into_paragraphs_newline(text)* | new line separator (`\n\n`) |

**Experiments**

Following the development of the chatbot and the implementation of various chunking strategies, instructional materials from two academic courses—Family Psychology and Sports Medicine—were integrated into the system to initiate the experimental phase. Each domain was evaluated independently to identify potential variations in chatbot performance across distinct subject areas.

To assess the accuracy of the chatbot's responses relative to the course content, the human expert and in the same time the educator of both courses formulated five (5) domain-relevant questions, which were employed as prompt queries. For each question, the chatbot generated two distinct response versions. These responses were subsequently evaluated by the respective human experts using a five-point Likert scale, ranging from 1 (Poor) to 5 (Excellent), based on predefined assessment criteria.

In total, six (6) distinct clustering models, as previously described, were applied to each question, yielding twelve (12) chatbot responses per question. To ensure comparability, the same set of clustering strategies was employed consistently across both domains, with retrieval constrained to the Top-K = 20 most relevant chunks for each question. Given the larger corpus size and greater conceptual complexity of the Sports Medicine domain, additional experiments were conducted using an extended retrieval depth of Top-K = 50 to investigate the relationship between retrieval breadth and chunking performance.

*Evaluation Criteria and Scoring*

Based on the five-point Likert evaluation scale applied to each assessment criterion, the minimum possible total score for a response version across the three criteria was three (3), representing a poor response, whereas the maximum attainable score was fifteen (15), representing an excellent response. A response was classified as satisfactory if it achieved a cumulative score exceeding eight (8) points across all criteria. According to the expert's evaluation, performance varied notably across the different chunking models.

Moreover, discernible differences were also identified between the two response versions generated within the same model. The mean of the two scores was calculated and used for subsequent analysis. The results were then compared across models, questions, and evaluation criteria, with all three criteria assigned equal weighting. Descriptive statistical analysis was conducted using mean scores and visualizations to summarize the numerical data.

*Domain-Specific Experimental Setup*

The *Family Psychology* domain comprised a textbook consisting of four chapters of educational material in Greek, corresponding to the content taught in the respective university course. The total size of the associated PDF files was approximately 1.65 MB. In contrast, the *Sports Medicine* domain exhibited a substantially larger and denser dataset, with the total size of the Greek-language PDF files exceeding 6.5 MB.

The use of the Greek language introduces additional linguistic challenges. Specifically, *Sports Medicine* terminology is characterized by flexible word order, extensive morphological variation, and considerable variability in sentence and paragraph length. These linguistic properties result in a more diffuse distribution of relevant information across the text, thereby reducing the efficiency of shallow retrieval methods. Given these characteristics, the *Sports Medicine* domain was selected to examine the relationship between retrieval depth *(Top-K)* and document separation strategies. Accordingly, experimental evaluations in this domain were conducted under two retrieval configurations: *Top-K* = 20 and *Top-K* = 50.

**Results**

*Answer to Research Question 1: For each model, which of the three evaluation criteria (content accuracy, completeness, clarity) represents its strongest aspect?*

The results obtained from the two courses reveal notable variations in performance. In the case of the *Family Psychology* course, the analysis demonstrated significant differences among the six (6) chunking models across the three evaluation criteria, as well as distinct patterns of performance at the question level. Specifically, based on the overall average scores per model and per question, the findings indicated that each model exhibited unique performance characteristics across the evaluated criteria.
Among the three criteria, **Content Accuracy** achieved the highest average performance, followed by **Clarity** and **Completeness**, respectively. Moreover, **Model 5** (*split_text_into_paragraphs_using_sentences(text)*) consistently outperformed the others across all criteria, with **Models 4** (*max_len parameter value* = 800) and **Model 6** (*split_text_into_paragraphs_newline(text)*) ranking next in performance. As illustrated in Figure 1, **Model 5** achieved the highest overall mean score (4.13), thereby demonstrating its superiority as the most effective chunking strategy within this experimental context.

## The total average score for all questions

| Model 1 | Model 2 | Model 3 | Model 4 | **Model 5** | Model 6 |
|---------|---------|---------|---------|-------------|---------|
| 3.60    | 3.63    | 3.70    | 4.10    | **4.13**    | 3.77    |

Figure 1. Overall performance of chunking models in Family Psychology (Top K = 20)

To conduct a more detailed assessment of performance across evaluation criteria, notable variations were observed among the models. For **Content Accuracy**, **Model 5** achieved the highest mean score (4.40), demonstrating superior capability in generating factually accurate and contextually relevant responses. **Models 1** and **4** also exhibited strong performance in this criterion. With respect to **Completeness**, **Model 4**—configured with *fixed-length* segmentation using *800*-character chunks—obtained the highest score (3.90). This outcome suggests that employing larger chunk sizes facilitated more comprehensive coverage of the input questions. **Model 5** followed closely with a score of 3.70. In terms of **Clarity**, **Model 5** again outperformed the other configurations (4.30), with **Model 4** showing comparable results (4.20). These findings indicate that both segmentation strategies contributed to the generation of coherent and well-structured responses.

To provide a comparative overview, Figure 1 illustrates the average scores for **Content Accuracy, Completeness,** and **Clarity** across all models within the *Family Psychology* domain.

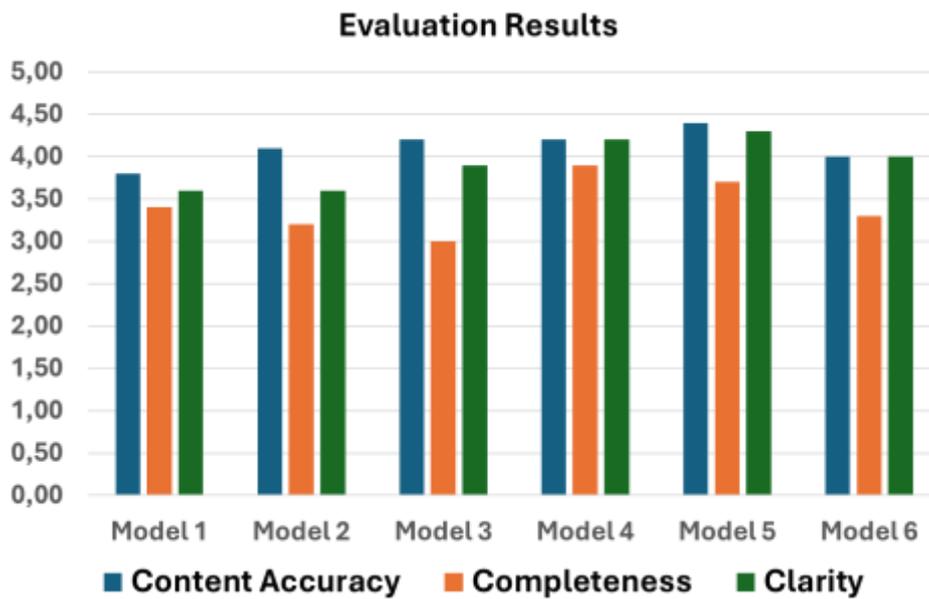

Figure 2. Average scores of the six chunking models across Content Accuracy, Completeness, and Clarity in the Family Psychology domain (Top K = 20).

As illustrated in Figure 2, certain models demonstrated clear strengths in specific criteria, which is further examined in the following section addressing Research Question 2.

Having established the performance patterns in the *Family Psychology* domain, the analysis was extended to the *Sports Medicine* domain to assess whether these patterns generalized in a more complex dataset. Following the methodology described above, we evaluated the performance of all six (2) chunking models in the *Sports Medicine* domain under two retrieval settings (*Top K = 20* and *Top K = 50*).

At *Top K = 20*, performance differences across models were modest, with a slight lead by *fixed-length, sentence-preserving* chunking (**Model 2**; average score 3.03), and **Clarity** tended to be the strongest criterion overall. This pattern suggests that shallow retrieval is sufficient to produce clear but not necessarily fully accurate or complete answers when dealing with complex technical content. Increasing retrieval depth to *Top K = 50* produced a clear improvement in overall performance. The *newline-paragraph* model (**Model 6**) became dominant (average score 4.00). As shown in Figure 3, **Model 2** achieved the highest average score at *Top K = 20*, while **Model 6** performed best at *Top K = 50*.

The total average score for all questions

| K = 20 | Model 1 | Model 2 | Model 3 | Model 4 | Model 5 | Model 6 |
|---|---|---|---|---|---|---|
| | 2.87 | 3.03 | 2.50 | 2.67 | 2.93 | 2.97 |

| K = 50 | Model 1 | Model 2 | Model 3 | Model 4 | Model 5 | Model 6 |
|---|---|---|---|---|---|---|
| | 3.57 | 3.30 | 3.50 | 3.47 | 3.17 | 4.00 |

Figure 3. Total average scores of the six chunking models in the Sports Medicine domain at two retrieval depths (Top K = 20 and Top K = 50), showing that Model 2 performed best at K = 20 and Model 6 at K = 50.

Based on the criterion-level analysis at *Top K = 20*:
- **Clarity** consistently outperformed the other two criteria across all models.
- In terms of **Content Accuracy**, **Model 2** achieved the highest score, followed closely by **Models 1** and **6**, while **Model 3** demonstrated the lowest accuracy.
- For **Completeness**, scores were generally modest, with **Model 5** performing slightly better than the other models, whereas **Model 3** received the lowest completeness score.

**Clarity** emerged as the strongest dimension overall, with **Model 5** obtaining the highest score, followed by **Model 4** and **Model 6**. This indicates that at shallow retrieval depth, most chunking strategies favored clear and well-structured responses rather than fully accurate or comprehensive ones. To provide a visual comparison of model performance across the three evaluation criteria at shallow retrieval depth, Figure 4 presents the average scores for **Content Accuracy, Completeness,** and **Clarity** at *Top K = 20*.

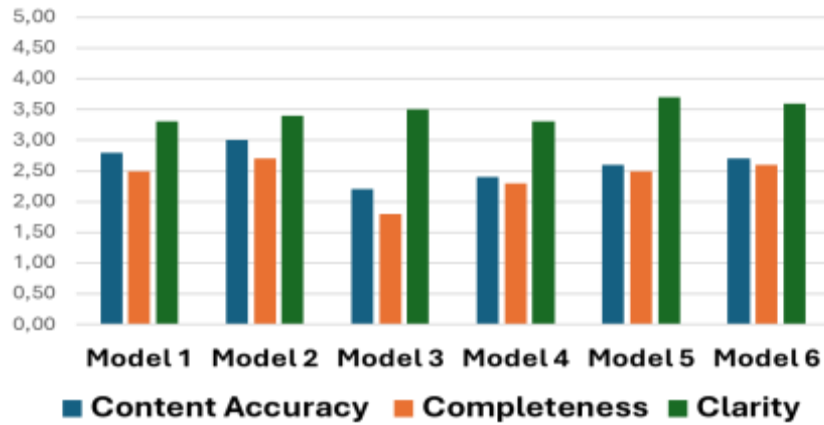

Figure 4. Average scores of the six chunking models across Content Accuracy, Completeness, and Clarity in the Sports Medicine domain (Top K = 20).

When the retrieval depth increased to *Top K = 50*, performance improved across all criteria. **Content Accuracy** showed a notable increase, with **Model 6** achieving the highest score (4.50), significantly outperforming the other models, followed by **Model 1** and **Model 3**. **Completeness** also improved, with **Model 6** and **Model 1** tying for the highest score, while **Model 4** remained the lowest. **Clarity** continued to be a strong criterion for most models, with **Model 5** achieving the highest clarity score, closely followed by **Models 6** and **2**. Unlike the *Top K = 20* condition, however, **Model 6** slightly surpassed **Clarity** in **Content Accuracy**, indicating that deeper retrieval enabled this model to integrate fine-grained evidence more effectively. These results highlight that increasing *Top K* amplifies the benefits of *paragraph-based* chunking strategies, leading to more accurate and informative responses. To examine the effect of deeper retrieval on model performance, Figure 5 illustrates the criterion-level scores at *Top K = 50*.

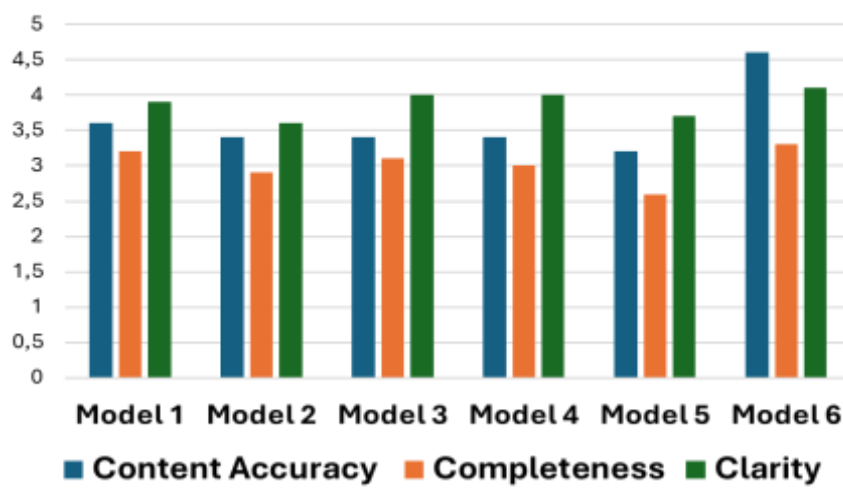

Figure 5. Average scores of the six chunking models across Content Accuracy, Completeness, and Clarity in the Sports Medicine domain (Top K = 50).

*Answer to Research Question 2: For each model, which of the three evaluation criteria (content accuracy, completeness, clarity) represents its strongest aspect?*

Regarding the strongest evaluation criterion for each chunking model in the *Family Psychology* course, the results showed a consistent pattern in favor of **Content Accuracy**. Specifically, **Models 1, 2,** and **3** demonstrated their highest performance in **Content Accuracy**. **Model 5** also achieved its strongest scores in **Content Accuracy**, with only a marginal difference from **Clarity**, while **Model 6** showed a nearly equal performance between **Content Accuracy** and **Clarity** but remained slightly higher in **Content Accuracy**. The only exception was **Model 4**, which performed best in **Clarity**, indicating that this chunking strategy facilitated more readable and well-structured responses even if **Accuracy** or **Completeness** were slightly lower.

Overall, five out of the six models (**Models 1, 2, 3, 5,** and **6**) showed **Content Accuracy** as their strongest criterion, suggesting that most chunking strategies supported accurate retrieval and generation more than **Clarity** or **Completeness**, while **Model 4** uniquely favored **Clarity**.

For the *Sports Medicine* course, the strongest evaluation criterion for each chunking model showed a highly consistent pattern in favor of **Clarity**. Specifically, **Models 1, 2, 3, 4** and **5** achieved their highest scores in **Clarity** at both retrieval depths (*Top K = 20* and *Top K = 50*), indicating that these chunking strategies supported well-structured and understandable responses even in a technical domain. **Model 6** exhibited a different behavior: at *Top K = 20*, its strongest criterion was also **Clarity** and at *Top K = 50,* **Content Accuracy** slightly surpassed **Clarity**. This suggests that deeper retrieval enabled this model to access and integrate complementary, fine-grained evidence distributed across multiple chunks, thereby producing more accurate responses.

Overall, for almost all models, **Clarity** was the strongest criterion, highlighting the importance of coherence and readability in technical content. Only **Model 6** at *Top K = 50* performed best in **Content Accuracy**, demonstrating that deeper retrieval can enhance accuracy when combined with paragraph-level chunking likely due to improved coverage of relevant information.

**Conclusion**

This study examined how different chunking strategies and retrieval depths influence the performance of a Greek-language RAG-based chatbot in higher education. The results clearly showed that response quality is not determined by a single model, but by the interaction between chunking and *Top K*. *Sentence-based* paragraphing (**Model 5**) worked best in expository content such as *Family Psychology*, while *paragraph-preserving* chunking (**Model 6**) combined with deeper retrieval (*Top K = 50*) achieved the highest accuracy in the denser and more technical *Sports Medicine* corpus. Based on a large corpus of Greek sports medicine texts in structured book format, the study shows that higher *Top K* values have a clear positive effect in domains with rich and complex content. Overall, the results show that chunking strategy and the choice of *Top K* are key factors for the performance of RAG-based chatbots, indicating that there is no universal configuration and that domain-aware optimization is essential.

From an educational perspective, the chatbot provided accurate, clear, and context-grounded responses aligned with course material, demonstrating its value as both a student support tool and an aid for educators in generating learning resources. Importantly, the study confirms that

RAG-based chatbots can operate effectively in linguistically complex languages such as Greek when supported by appropriate retrieval and segmentation strategies, contributing to more inclusive AI-enhanced learning environments.

Future work could explore adaptive retrieval mechanisms, integration with knowledge graphs, and real-world classroom deployment to evaluate long-term learning impact.
In general, findings support the potential of RAG-based chatbots to transform static educational content into interactive, pedagogically meaningful support. Future research will focus on testing the system in real higher education settings with both students and educators, as well as expanding the evaluation to additional domains. An important direction is to examine how different chunking strategies can be tailored to the characteristics of Greek-language academic texts, bringing theory and practice closer together. Taken together, these findings support the potential of RAG-based chatbots to transform static educational content into dynamic, pedagogically meaningful interaction in higher education.

**Declaration of Generative AI and AI-assisted Technologies in the Writing Process**

The AI-assistive technology, ChatGPT, has been used for text improvement and proofreading the manuscript.

**Contact e-mail:** mariloukoutsi@gmail.com